\documentclass[aps, prl, superscriptaddress, twocolumn, preprintnumbers,amsmath,amssymb,notitlepage, nobalancelastpage,10pt,longbibliography]{revtex4-2}

\usepackage{amsmath}
\usepackage{verbatim}
\usepackage{graphicx}
\usepackage{color}
\usepackage{braket}
\usepackage{yfonts}
\usepackage{soul}
\usepackage[normalem]{ulem}
\usepackage{dsfont}

\newcommand{\cop}{\hat{c}}
\newcommand{\cdop}{\hat{c}^\dag}
\newcommand{\Hop}{\hat{H}}
\newcommand{\nop}{\hat{n}}
\newcommand{\llangle}{\left\langle\!\left\langle}
\newcommand{\rrangle}{\right\rangle\!\right\rangle}
\newcommand{\lllangle}{\left\langle\!\left\langle\!\left\langle}
\newcommand{\rrrangle}{\right\rangle\!\right\rangle\!\right\rangle}

\renewcommand{\Im}{\text{Im}}

\newcommand{\mean}[1]{\left\langle{#1}\right\rangle}

\usepackage[colorlinks=true,citecolor=blue,linkcolor=blue,urlcolor=blue]{hyperref}
\usepackage{subfigure}

\begin{document}

\title{Topological stripe state in an extended Fermi-Hubbard model}

\author{Sergi Juli\`a-Farr\'e}
\email{sergi.julia@icfo.eu}
\affiliation{ICFO - Institut de Ciencies Fotoniques, The Barcelona Institute of Science and Technology, Av. Carl Friedrich Gauss 3, 08860 Castelldefels (Barcelona), Spain}
\author{Lorenzo Cardarelli}
\affiliation{Peter Gr\"unberg Institute, Theoretical Nanoelectronics, Forschungszentrum J\"ulich, D-52428 J\"ulich, Germany}
\affiliation{Institute for Quantum Information, RWTH Aachen University, D-52056 Aachen, Germany}
\author{Maciej Lewenstein}
\affiliation{ICFO - Institut de Ciencies Fotoniques, The Barcelona Institute of Science and Technology, Av. Carl Friedrich Gauss 3, 08860 Castelldefels (Barcelona), Spain}
\affiliation{ICREA, Pg. Llu\'is Companys 23, 08010 Barcelona, Spain}
\author{Markus M\"uller}
\affiliation{Peter Gr\"unberg Institute, Theoretical Nanoelectronics, Forschungszentrum J\"ulich, D-52428 J\"ulich, Germany}
\affiliation{Institute for Quantum Information, RWTH Aachen University, D-52056 Aachen, Germany}
\author{Alexandre Dauphin}
\email{alexandre.dauphin@icfo.eu}
\affiliation{ICFO - Institut de Ciencies Fotoniques, The Barcelona Institute of Science and Technology, Av. Carl Friedrich Gauss 3, 08860 Castelldefels (Barcelona), Spain}

\begin{abstract}
Interaction-induced topological systems have attracted a growing interest for their exotic properties going beyond the single-particle picture of topological insulators. In particular, the interplay between strong correlations and finite doping can give rise to nonhomogeneous solutions that break the translational symmetry. In this work, we report the appearance of a topological stripe state in an interaction-induced Chern insulator around half-filling. In contrast to similar stripe phases in nontopological systems, here we observe the appearance of chiral edge states on top of the domain wall. Furthermore, we characterize their topological nature by analyzing the quantized transferred charge of the domains in a pumping scheme. Finally, we focus on aspects relevant to observing such phases in state-of-the-art quantum simulators of ultracold atoms in optical lattices. In particular, we propose an adiabatic state preparation protocol and a detection scheme of the topology of the system in real space.
\end{abstract}

\maketitle
{\textit{Introduction}}.\textemdash In the last decade, the quest for materials exhibiting intrinsic topological phases in the absence of external fields has been the focus of very intense research~\cite{Rachel_2018,liu_2016_reviewqah,Laughlin_csl_1990}. The interaction-induced  quantum anomalous Hall (QAH) phase~\cite{liu_2016_reviewqah,Chang_2013_dopingqah,Chang2015_qahprl,Deng2020,Serlin2020,Li_2021_tmd}, or chiral spin liquids~\cite{WenWilczekZee_csl_1989,Laughlin_csl_1990,semeghini_probing_2021}, are two paradigmatic examples. In both cases, the interplay between the interactions and the geometry leads to the spontaneous breaking of time-reversal symmetry, and the resulting phases possess nontrivial topological invariants. The theoretical search of such interaction-induced topological phases in many-body systems has been further boosted by the development of tensor network approaches~\cite{Orus_2019_review,schollwock_density-matrix_2011}. In particular, state-of-the-art density matrix renormalization group (DMRG) studies~\cite{Stoudenmire2012} in cylinder geometries have unambiguously established the presence of spontaneous Chern insulators in the ground-state phase diagram of several two-dimensional lattice models. These include effective models of twisted bilayer graphene~\cite{Chen_2021_TBG}, or extended  Fermi-Hubbard models of spinless fermions ~\cite{zhu_interaction-driven_2016,Sur2018,zeng_tuning_2018} that can be engineered in cold atom quantum simulators. Furthermore, fractional Chern insulators have been also identified in the spinful Fermi-Hubbard model~\cite{Szasz2020} and in the Heisenberg model~\cite{Gong_2014_pumpdmrg}, both representing cases in which the system realizes a chiral spin liquid phase.

While all these studies focused on spatially homogeneous phases at commensurate particle fillings, it is worth noticing that the study of inhomogeneous phases at incommensurate fillings, i.e., at finite doping, is of particular interest. Several works in this direction have pushed tensor network simulations to their limit in order to identify antiferromagnetic stripe domain walls of high-$T_c$ superconductors in the underdoped region of the Hubbard model~\cite{Huang2017,Zheng2017}, as first predicted by mean-field studies~\cite{Machida89,Schulz89}. In the case of interaction-induced Chern insulating phases, very recent mean-field studies~\cite{julia-farre_self-trapped_2020,Shi2021,Kwan2021,Shin2021} suggested that at incommensurate dopings these systems can also exhibit domain walls between phases characterized by different topological invariants, leading to interaction-induced chiral edge states~\cite{julia-farre_self-trapped_2020}. Remarkably, this picture is consistent with the subsequent experimental observation of a mosaic of patches with opposite topological invariants in 
twisted bilayer graphene~\cite{Grover2021_mosaicChern}. 

In this work, we analyze the phenomenon of spatially inhomogeneous topological phases in 2D. Based on a DMRG study in the matrix-product-state (MPS) representation, we confirm the numerical stability of these phases beyond the mean-field approximation in a cylinder geometry with a very long length and short transverse direction. We also  introduce techniques to measure topological invariants in inhomogeneous systems and in a purely many-body scenario, i.e., beyond the single-particle approximation. 

To this aim, we consider the effect of doping in the interaction-induced homogeneous QAH phase of a fermionic lattice model. We start by showing that such a system indeed exhibits a topological stripe state, hosting chiral edge states at the domain walls. We then characterize the topological nature of the domains by means of a topological pumping scheme. Following Laughlin's Gedankenexperiment~\cite{Laughlin1981_pump}, which we generalize to the inhomogeneous case, we extract the Chern number of the domains from their quantized charge transfer under an adiabatic flux insertion in the DMRG simulations.

Our study not only reveals the fundamental features of these inhomogeneous solutions, and how they can be characterized in a strongly-correlated scenario. It is also further motivated by the prospect of quantum simulating these phases with cold atoms in optical lattices. In this regard, notice that in solid-state materials the QAH has only been observed in a few systems with spin-orbit coupling~\cite{Chang_2013_dopingqah, Chang2015_qahprl,Deng2020} or with interacting magnetic orbitals~\cite{Serlin2020,Li_2021_tmd}. On the other hand, noninteracting Chern insulators have also been observed in quantum simulators~\cite{Jotzu_2014,Mancini2015,Asteria_2018,Wintersperger2020} via the engineering of artificial gauge fields~\cite{Goldman_2014,Celi_2014}. The extension of these experiments to the interacting case would allow one to observe new phenomena. Motivated by these reasons, we propose schemes to prepare these phases in an experiment and develop strategies to characterize their topology in real space. In this context, we show that the topological phase of the model could be prepared in a quasi-adiabatic protocol via a control parameter of the lattice that induces a continuous topological phase transition. Such a result is essential to provide a path to  adiabatically prepare the phase in an experimental setup. We finally discuss the possibility of measuring the topological nature of the phase through snapshot measurements of the particle density. 

\begin{figure}[t]
    \centering
    \includegraphics[width=\columnwidth]{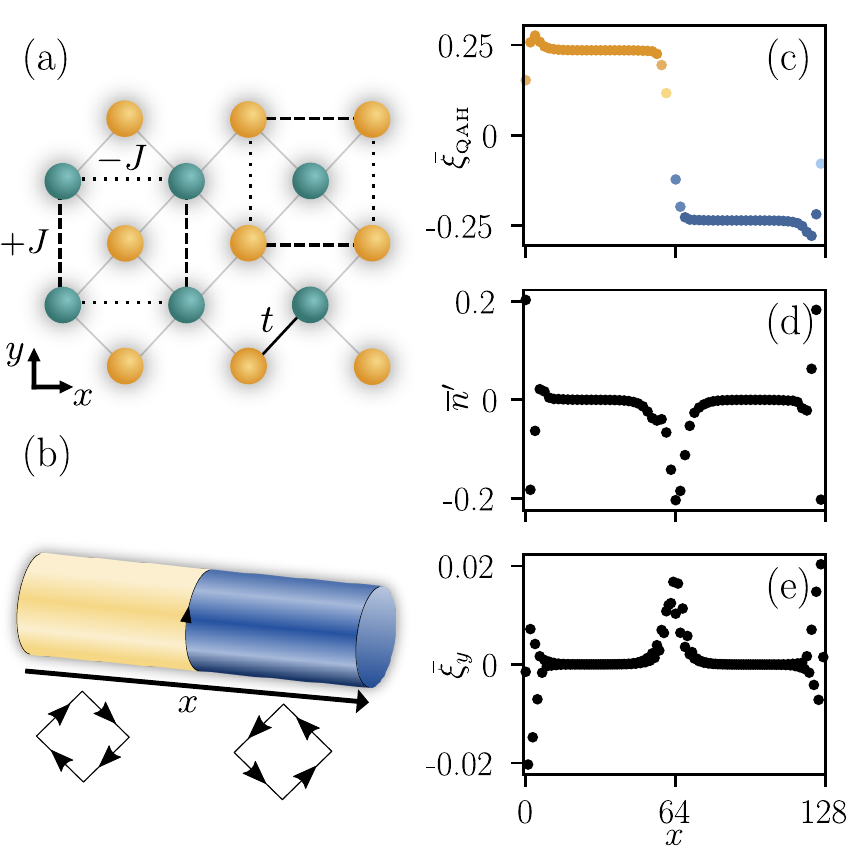}
    \caption{(a) Hopping processes of the Hamiltonian on the checkerboard lattice. (b) Sketch of the topological stripe state in the cylinder geometry. (c)-(e) Expectation value of local quantities integrated over the radial direction of the cylinder. (c) Current loop order featuring a sign inversion at the center of the cylinder. (d) Deviation of the local density from half filling. One can clearly observe the presence of the hole at the center. (e) Radial currents signaling the presence of chiral edge states in the regions where there is a change in the Chern number.}
    \label{fig:local_orders}
\end{figure}

\textit{Model}.\textemdash We consider the extended Fermi-Hubbard Hamiltonian of spinless fermions on a checkerboard lattice described by the Hamiltonian $\Hop=\Hop_0+\Hop_\text{int}$. The quadratic part $\Hop_0$  of the Hamiltonian reads
\begin{equation} \label{eq:hamiltonianfree}
\begin{split}
\Hop_{0} = &-t\sum_{\langle ij\rangle}(\cdop_{i}\cop_{j}+\textrm{H.c.})+J\sum_{\llangle ij\rrangle}e^{i\phi_{ij}}(\cdop_{i}\cop_{j}+\textrm{H.c.}),
\end{split}
\end{equation}
where $t$ and $J$ are the nearest-neighbor (NN) and next-to-nearest-neighbor (NNN) hopping amplitudes, respectively [see Fig.~\ref{fig:local_orders}(a)]. The phase $\phi_{ij}=\pm \pi$ of the NNN tunneling generates a $\pi$-flux on each sublattice. On the other hand, the interacting part $\Hop_\text{int}$ of the Hamiltonian has repulsive density-density interaction up to third neighbors and reads
\begin{equation} \label{eq:hamiltonianinter_p}
\begin{split}
\Hop_\textrm{int}= V_1\sum_{\langle ij\rangle} \hat{n}'_{i}\hat{n}'_{j}+V_2\sum_{\llangle ij\rrangle}\hat{n}'_{i}\hat{n}'_{j}
+V_3\sum_{\lllangle ij \rrrangle}\hat{n}'_{i}\hat{n}'_{j},
\end{split}
\end{equation}
with $\hat{n}'_{i}\equiv \hat{n}_{i}-1/2$ and $\hat{n}_i=\hat{c}^\dagger_i \hat{c}_i$. 
At half filling, $\hat{H}_0$ exhibits two bands with a quadratic band touching (semi-metallic phase).  For finite interactions $V_1/2\simeq V_2 \gg V_3$, the frustration induced by the competition between semi-classical charge orders allows for the emergence of an interaction-induced QAH state in the phase diagram~\cite{Sur2018,julia-farre_self-trapped_2020,cardarelli2022}. The latter is characterized by the appearance of spatially homogeneous local current loop order, $\xi_{\textrm{QAH}}\equiv \sum_{ij \in \textrm{plaq.}} \Im\,\mean{\hat{c}^\dagger_i\hat{c}_j}$, in NN plaquettes (see \cite{SM} for details), which breaks time-reversal symmetry spontaneously. In addition, it is also characterized by a nonzero value of a global topological invariant, the many-body Chern number $\nu$. Importantly, there is an exact twofold ground state degeneracy, corresponding to the two opposite values of $\xi_\textrm{QAH}$. These two sectors are therefore characterized by opposite Chern numbers $\nu_\pm=\pm 1$.

\textit{Topological stripe state}.\textemdash We now discuss the appearance of spatially inhomogeneous Chern insulators in the model around half-filling, which constitutes one of the central results of this work. We consider a cylinder geometry with 6 two-site unit cells in the radial direction ($y$) and 64 in the longitudinal one ($x$). To determine its ground state, we use the DMRG algorithm on the one-dimensional folding of the cylinder. In the numerical treatment, this 1D system, therefore, has effective long-range Hamiltonian terms, and one needs to use large bond dimensions $\chi_\textrm{max}=3000$ in order to get truncation errors of the order $10^{-5}$ at most. At half filling, for $V_1/t=4.5,\ V_2/t=2.25, V_3/t=0.5$ and $J/t=0.5$, the system presents a degenerate QAH ground state with Chern numbers $\nu_\pm=\pm1$. The addition of a single hole favors the breaking of the translational symmetry, as shown in Fig.~\ref{fig:local_orders}. 
Figure~\ref{fig:local_orders}(b) depicts the DMRG solution, which we call the topological stripe state. Such a state is spatially composed of two different Chern insulators, located on distinct halves of the cylinder and separated by a stripe domain wall. That is, due to the spontaneous breaking of translational invariance induced by doping, the two degenerate ground states of half-filling coexist in two separate regions of the same bulk. 
Figure~\ref{fig:local_orders}(d) shows the density profile integrated along the radial direction. We observe that the domain wall is induced by the presence of a hole-like stripe, which is located in the bulk of the cylinder and has an integrated quantized charge of $Q=-1$~\footnote{We also observe density and current fluctuations at the edges. However, they do not have net charge. Furthemore, these edge currents switch sign and are thefore not chiral. They are therefore not associated to topological edge states.}. The latter separates two different Chern insulators, as signaled by the inversion of the current loop order $\xi_\textrm{QAH}$, shown in Fig.~\ref{fig:local_orders}(c). Notice that this is reminiscent of the change in the phase of the antiferromagnetic order parameter observed in the stripe phase of the Fermi-Hubbard mode, in the context of cuprate high-$T_\textrm{c}$ superconductors~\cite{Huang2017,Zheng2017}. 
Here, however, the local order parameter $\xi_\textrm{QAH}$ is intertwined with the topological Chern number $\nu$. This enriches the features of this topological stripe state, compared to the case of nontopological magnetic stripes. For instance, by virtue of the bulk-edge correspondence of topological insulators, one expects the presence of chiral edge states at the interface between the two different Chern insulators. Furthermore, these chiral edge states should have chiral current in the radial direction, defined as ${\xi}_{y}^{ij}\equiv 2Je^{i\phi_{ij}}\langle \cdop_i\cop_j\rangle$, where $(i,j)$ are NNN bonds in the radial direction. This quantity integrated in the radial direction is shown in Fig.~\ref{fig:local_orders}(e). We observe positive net currents around the position of the hole, where the topological invariant changes its value, as discussed below.  
\begin{figure}[t]
    \centering
    \includegraphics[width=1\columnwidth]{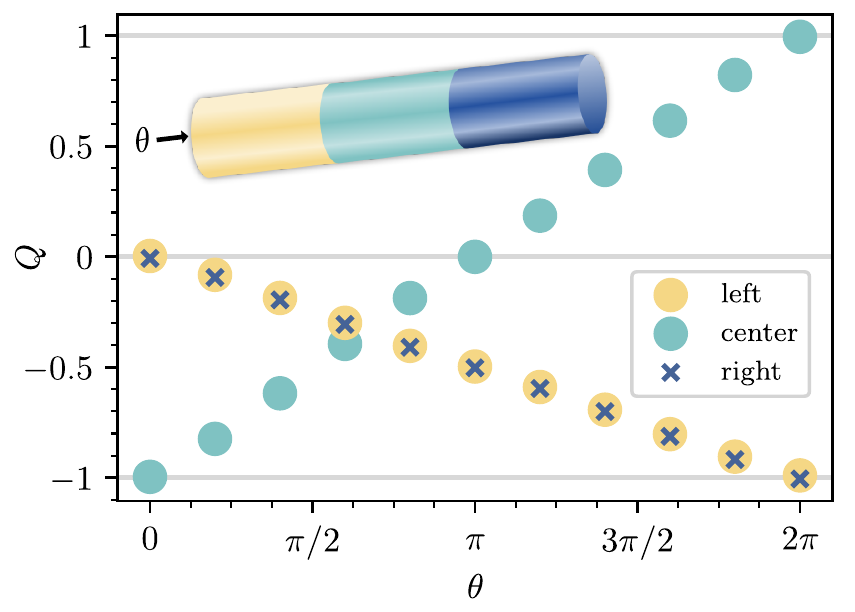}
    \caption{Quantized charge transport in the topological pump procedure performed with an adiabatic DMRG simulation. The net charges of the left, central, and right regions are shown in yellow, green, and blue colors, respectively, and as a function of the inserted flux $\theta$, as indicated in the inset sketch.}
    \label{fig:pump}
\end{figure}

\textit{Topological pump in inhomogeneous Chern insulators}.\textemdash While the local quantities shown in Figs.~\ref{fig:local_orders}(b)-(e) are consistent with a topological stripe state, where each of the sides of the cylinder has a different Chern number, one needs to explicitly compute these global invariants in order to rigorously characterize the topological nature of this state. Notice that, for such an interacting and inhomogeneous state, this task is particularly challenging, as the main tools to study topology in real space, e.g., the local Chern marker~\cite{BiancoThesis,PhysRevB.84.241106}, are limited to the free fermionic picture, where interactions can only be treated in mean-field approximation~\cite{Irsigler2019,julia-farre_self-trapped_2020}. 
Here, to compute a spatially inhomogeneous Chern number in a purely many-body scenario, we follow the adiabatic flux insertion procedure, introduced as a \textit{gedankenexperiment} by Laughlin~\cite{Laughlin1981_pump}. This is based on the fact that Chern insulators exhibit a quantized Hall response equal to the Chern number after one cycle of a charge pump. While this method has been widely used in adiabatic DMRG simulations of homogeneous systems to compute their integer~\cite{zhu_interaction-driven_2016,Sur2018} and fractional~\cite{Gong_2014_pumpdmrg,Szasz2020} Chern numbers, here we show that it is also suited to analyze the topology of inhomogeneous stripe states.
We insert a $U(1)$ flux to the stripe ground state obtained in the previous section by adiabatically changing the phase of the tunneling terms crossing the $y$ periodic boundary $\hat{c}^\dagger_i\hat{c}_j \rightarrow \hat{c}^\dagger_i\hat{c}_j e^{i \theta} $ in the DMRG simulation~\footnote{In the adiabatic flux insertion, we obtain the state $\ket{\Psi(\theta)}$ by initializing the DMRG algorithm in the previous converged state $\ket{\Psi(\theta-\delta\theta)}.$ For sufficiently small $\delta\theta$, this is equivalent to the dynamics generated by a slowly varying time-dependent flux Hamiltonian $\hat{H}[\theta(t)]$. Importantly,  the finite bond dimension and the local update of the DMRG algorithm ensure the adiabatic condition $\lim_{\delta\theta\to 0}\braket{\Psi(\theta)|\Psi(\theta-\delta\theta)}=1$. In particular, it avoids crossings between topological edge modes inside the insulating gap during the flux insertion, which would lead to quantized jumps in the pumped charge~\cite{Hatsugai2016,SM}.}. For a full cycle $\theta:0\rightarrow 2\pi$, and according to Laughlin's argument, a homogeneous Chern insulator in a cylinder geometry pumps a quantized charge $\Delta Q$ equal to the value of $\vert\nu\vert$ from left to right, or vice versa, depending on the sign of the Chern number. 
For an inhomogeneous system with two different nontrivial Chern numbers, we instead expect a quantized transport from the edges to the center, or vice versa, as discussed below. The effect of the flux insertion in the topological stripe state can be seen in Fig.~\ref{fig:pump}, which shows the evolution of the integrated charge deviation from half-filling, defined as $Q_\mathcal{S,\theta}\equiv \sum_{i \in \mathcal{S}}\hat{n}'_i(\theta)$, where $\mathcal{S}\in\lbrace {l,c,r}\rbrace$ corresponds to the left, center, or right region of the cylinder, respectively. We also define the transferred charge on each region during the pump as $\Delta Q_\mathcal{S}\equiv Q_{\mathcal{S},2\pi}-Q_{\mathcal{S},0}$. At the beginning of the pump, $Q_{l,0}=Q_{r,0}=0$, and $Q_{c,0}=-1$, as the added hole is located in the central region. As $\theta$ increases, the combination of the Hall responses on each half of the cylinder leads to a net accumulation of charge in the domain wall, which indicates that the two halves of the cylinder have different Chern numbers. That is, for a unique value of the Chern number the charge would instead flow from one edge to the other without a net accumulation in the bulk. Indeed, notice that the charge pumped to the center domain wall is related to the Chern numbers of the left and right halves of the cylinder through
\begin{equation}\label{eq:inhomogeneous_chern}
    \Delta Q_{c}\equiv -(\Delta Q_{l}+\Delta Q_{r})=\nu_{l}-\nu_{r}.
\end{equation}
At the end of the cycle ($\theta=2\pi$), we observe that both the left and right halves have transported a unit charge to the center, and the initial central hole is converted into a particle, i.e., $\Delta Q_\textrm{c}=2$. This is in agreement with these two regions having different Chern numbers $\nu_l=1$ and $\nu_r=-1$. Therefore, with the help of Eq.~\eqref{eq:inhomogeneous_chern} and the DMRG adiabatic flux insertion, we are able to unambiguously establish the topological character of this spatially inhomogeneous phase. For completeness, we also provide a qualitative single-particle explanation of this generalized Laughlin pump for inhomogeneous systems in the Supplemental Materials~\cite{SM}.

\begin{figure}[t]
    \centering
    \includegraphics[width=1\columnwidth]{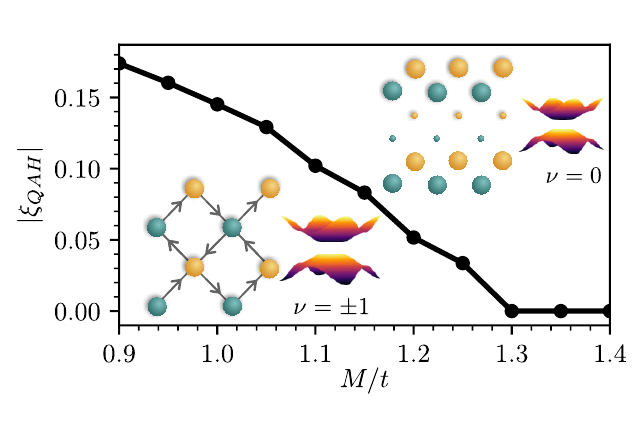}
    \caption{Adiabatic state preparation of the interaction-induced QAH phase via the lattice control parameter $M$. The continuous behavior of the local order parameter $\xi_\textrm{QAH}$ indicates a continuous phase transition from the trivial stripe insulator at $M/t\gg 1$ to the QAH state at $M\to 0$.}
    \label{fig:state_prep}
\end{figure}

\textit{Adiabatic state preparation of the interaction-induced QAH phase}.\textemdash
Compared to other QAH states emerging from spontaneous symmetry breaking in solid-state systems, the one considered here is described by a relatively simple Hamiltonian  $\hat{H}$ that can be quantum-simulated in a controlled environment. In particular, Rydberg-dressed atoms in optical lattices can be used to simulate such an extended Fermi-Hubbard model with tunable long-range interactions~\cite{guardado-sanchez_quench_2021,cardarelli2022}. Here we focus on the yet unaddressed question of the quantum state preparation of this exotic phase, which is ultimately related to the appearance of the domain wall states discussed above. For the adiabatic state preparation of the QAH phase~\cite{Schutzhold2006_prep,Barkeshli2015_prep} it is desirable to find a second-order phase transition from a trivial insulator that could be easily initialized~\cite{Popp2004_prep,Sorensen2010}. This strategy has already been used to prepare noninteracting Chern insulators in optical lattices~\cite{aidelsburger_realization_2013}, and in the presence of interactions, there are numerical proposals to prepare fractional Chern insulators~\cite{Ye2017_prep,Motruk2017_prep}. The main difference in the present case is that the QAH phase arises from the spontaneous breaking of time-reversal symmetry in the ground state, that is, in the absence of external gauge fields. Therefore, we expect the appearance of Kibble-Zurek defects in a continuous transition~\cite{Kibble_1976,Zurek_1985,Zurek_2005,KibbleZurekRydberg}, qualitatively resembling the static stripe state discussed above, and their interplay with topological chiral edge states. 

For the Hamiltonian $\hat{H}$ under consideration, however, all the interaction-induced charge orders in the phase diagram feature a first-order phase transition to the QAH state~\cite{Sur2018}. To overcome this problem, we propose to add to $\hat{H}$ a staggering potential~\cite{Aidelsburger_2014,Motruk2017_prep} with strength $M$ of the form
\begin{equation}
    \hat{H}_\textrm{prep}=\frac{M}{2}\sum_i (-1)^{s_i}\nop_i,
\end{equation}
where $s_i=\pm 1$ on alternating two-site longitudinal stripes (see Fig.~\ref{fig:state_prep}), which in the absence of interactions induces a local charge order at half filling corresponding to alternating empty and occupied stripes. In order to analyze the nature of the phase transition when varying $M$, we use the infinite density-matrix-renormalization-group (iDMRG) in the cylinder geometry with a single ring unit cell. Compared to the previous finite DMRG simulation of the topological stripe state, here we need to enlarge the bond dimension to $\chi_\textrm{max}=4000$ to stabilize solutions with a small but finite value of the current loop order $\xi_{\textrm{QAH}}$. As shown in Fig.~\ref{fig:state_prep}, when $M$ dominates, the system is in a trivial charge insulating state with a vanishing current loop order $\xi_\textrm{QAH}$. Upon decreasing $M$, the local order parameter $\xi_\textrm{QAH}$ becomes finite without exhibiting a clear discontinuous jump, which suggests a continuous phase transition to the QAH phase.  

\begin{figure}[t]
    \centering
    \includegraphics[width=\columnwidth]{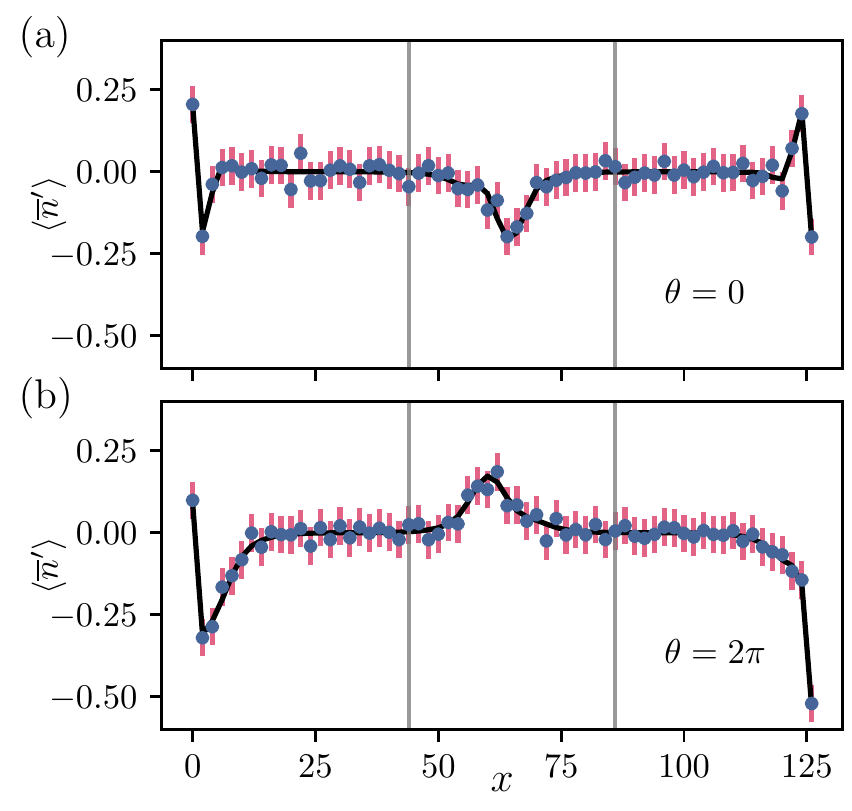}
    \caption{Computation of quantized Hall responses via local density snapshots in the topological pump procedure. (a),(b) Estimated density profiles from 3500 snapshots at the beginning and at the end of the flux insertion cycle, respectively. The Chern numbers of the left and right regions are extracted from the difference between these two cases.}
    \label{fig:snapshots}
\end{figure}

\textit{Snapshot-based detection of the Chern number in transport experiments with cold atoms}.\textemdash One of the advantages of the numerical determination of Chern numbers via the topological pump procedure described above is that it can be connected to the experimental measurement of this global topological invariant in real space. For instance, the 2D Laughlin topological pump itself has been experimentally realized for noninteracting particles with cold atom quantum simulators in a synthetic cylinder geometry~\cite{Fabre2022}. Moreover, in a 2D lattice with open boundary conditions, the presence of an external force playing the role of an electric field is expected to result in the same quantized Hall response~\cite{Motruk2020_quantizedtransport}. In both cases, the Chern number can be related to the charge drift in the system, which can be extracted from snapshots of the local density, accessible with a quantum gas microscope~\cite{Bakr2009,Weitenberg2011}. Here, to numerically simulate snapshot measurements at the initial and final stages of the topological pump, we use an algorithm proposed by Ferris and Vidal~\cite{FerrisVidal2012}. In a nutshell, this method allows one to efficiently draw independent snapshots of the local density of an MPS, by simulating collapse measurements in the occupation basis at each site. The results are shown in Fig.~\ref{fig:snapshots}, which shows the averaged values $\langle\bar{n}'\rangle$ for 3500 snapshots. In Fig.~\ref{fig:snapshots}(a), corresponding to the $\theta=0$ case, the central hole is signaled by the depletion of the local density in this region. In this case, the deviation charges on the left and right regions are estimated, respectively, as $Q_{l,0}=(0.01\pm 0.26)$ and $Q_{r,0}=(-0.01\pm 0.26)$. At the final stage of the pump [Fig.~\ref{fig:snapshots}(b)], one observes an excess charge in the central region, and the left and right regions have nonvanishing net charges of $Q_{l,2\pi}=(-0.95\pm 0.26)$ and $Q_{{r},2\pi}=(-0.97\pm 0.26)$, respectively. From these quantities, we estimate the Chern number of the left and right regions as $\nu_{l}=(0.96\pm 0.37)$ and $\nu_{r}=-(0.96\pm 0.37)$, which are compatible with the ones extracted from Fig.~\ref{fig:pump}.

\textit{Conclusions}.\textemdash We provided numerical evidence of a topological stripe state in an extended Fermi-Hubbard model at finite hole doping in a cylinder geometry. We generalized the numerical Laughlin pump procedure to characterize the two spatially separated Chern numbers of such a state. We furthermore discussed a related detection scheme on a quantum simulator based on snapshot measurements of the local density.
 Our methods can be easily adapted to analyze other interacting systems with inhomogeneous topological properties in real space.

\begin{acknowledgments}
\textit{Acknowledgments}.\textemdash ICFO group acknowledges support from: ERC AdG NOQIA; Ministerio de Ciencia y Innovation Agencia Estatal de Investigaciones (PGC2018-097027-B-I00/10.13039/501100011033, CEX2019-000910-S/10.13039/501100011033, Plan National FIDEUA PID2019-106901GB-I00, FPI (reference code BES-2017-082118), QUANTERA MAQS PCI2019-111828-2, QUANTERA DYNAMITE PCI2022-132919, Proyectos de I+D+I “Retos Colaboración” QUSPIN RTC2019-007196-7); MCIN Recovery, Transformation and Resilience Plan with funding from European Union NextGenerationEU (PRTR C17.I1); Fundació Cellex; Fundació Mir-Puig; Generalitat de Catalunya (European Social Fund FEDER and CERCA program (AGAUR Grant No. 2017 SGR 134, QuantumCAT \ U16-011424, co-funded by ERDF Operational Program of Catalonia 2014-2020); Barcelona Supercomputing Center MareNostrum (FI-2022-1-0042); EU Horizon 2020 FET-OPEN OPTOlogic (Grant No 899794); ICFO Internal “QuantumGaudi” project; EU Horizon Europe Program (Grant Agreement 101080086 — NeQST), National Science Centre, Poland (Symfonia Grant No. 2016/20/W/ST4/00314); European Union’s Horizon 2020 research and innovation program under the Marie-Skłodowska-Curie grant agreement No 101029393 (STREDCH) and No 847648 (“La Caixa” Junior Leaders fellowships ID100010434: LCF/BQ/PI19/11690013, LCF/BQ/PI20/11760031, LCF/BQ/PR20/11770012, LCF/BQ/PR21/11840013). Views and opinions expressed in this work are, however, those of the authors only and do not necessarily reflect those of the European Union, European Climate, Infrastructure and Environment Executive Agency (CINEA), nor any other granting authority. Neither the European Union nor any granting authority can be held responsible for them. The RWTH and FZJ group acknowledges support by the ERC Starting Grant QNets Grant Number 804247, the EU H2020-FETFLAG-2018-03 under Grant Agreement number 820495, by the Germany ministry of science and education (BMBF) via the VDI within the project IQuAn, by the Deutsche Forschungsgemeinschaft through Grant No. 449905436, and by US A.R.O. through Grant No. W911NF-21-1-0007, and by the Office of the Director of National Intelligence (ODNI), Intelligence Advanced Research Projects Activity (IARPA), via US ARO Grant number W911NF-16-1-0070. All statements of fact, opinions, or conclusions contained herein are those of the authors and should not be construed as representing the official views or policies of ODNI, the IARPA, or the US Government.
The authors gratefully acknowledge the computing time provided to them at the NHR Center NHR4CES at RWTH Aachen University (project number p0020074). This is funded by the Federal Ministry of Education and Research, and the state governments participating on the basis of the resolutions of the GWK for national high-performance computing at universities (www.nhr-verein.de/unsere-partner).
The finite DMRG and iDMRG) simulations were performed using the iTensor~\cite{itensor}  and TenPy~\cite{tenpy} libraries, respectively.
\end{acknowledgments}

\end{document}


\title{Supplemental Materials: Topological stripe state in an extended Fermi-Hubbard model}

\author{Sergi Juli\`a-Farr\'e}
\email{sergi.julia@icfo.eu}
\affiliation{ICFO - Institut de Ciencies Fotoniques, The Barcelona Institute of Science and Technology, Av. Carl Friedrich Gauss 3, 08860 Castelldefels (Barcelona), Spain}
\author{Lorenzo Cardarelli}
\affiliation{Peter Gr\"unberg Institute, Theoretical Nanoelectronics, Forschungszentrum J\"ulich, D-52428 J\"ulich, Germany}
\affiliation{Institute for Quantum Information, RWTH Aachen University, D-52056 Aachen, Germany}
\author{Maciej Lewenstein}
\affiliation{ICFO - Institut de Ciencies Fotoniques, The Barcelona Institute of Science and Technology, Av. Carl Friedrich Gauss 3, 08860 Castelldefels (Barcelona), Spain}
\affiliation{ICREA, Pg. Llu\'is Companys 23, 08010 Barcelona, Spain}
\author{Markus M\"uller}
\affiliation{Peter Gr\"unberg Institute, Theoretical Nanoelectronics, Forschungszentrum J\"ulich, D-52428 J\"ulich, Germany}
\affiliation{Institute for Quantum Information, RWTH Aachen University, D-52056 Aachen, Germany}
\author{Alexandre Dauphin}
\email{alexandre.dauphin@icfo.eu}
\affiliation{ICFO - Institut de Ciencies Fotoniques, The Barcelona Institute of Science and Technology, Av. Carl Friedrich Gauss 3, 08860 Castelldefels (Barcelona), Spain}

\renewcommand{\theequation}{S\arabic{equation}}
\renewcommand{\thesection}{S\Roman{section}}
\renewcommand{\thefigure}{S\arabic{figure}}
\renewcommand{\bibnumfmt}[1]{[S#1]}
\renewcommand{\citenumfont}[1]{S#1}

\maketitle

\section{Definition of the quantum anomalous Hall local order parameter}

\begin{figure}[h]
    \centering
    \includegraphics[width=0.2\columnwidth]{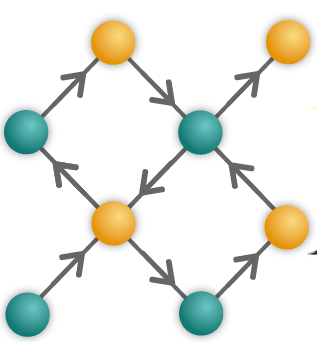}
    \caption{Depiction of the checkerboard lattice with the arrow convention for computing the current loop order parameter.}
    \label{fig:loop_order}
\end{figure}

In order to compute the local order parameter of the QAH phase $\xi_{\textrm{QAH}}=\sum_{ij\in \textrm{plaquette}}\textrm{Im}\mean{\hat{c}_i^\dagger\hat{c}_j}$, introduced in the main text, we need to follow the arrow convention shown in Fig.~\ref{fig:loop_order} for the bond $i\rightarrow j$ in each of the bulk plaquettes.

\section{Details on the DMRG calculations}
The DMRG (iDMRG) calculations are performed with the iTensor~\cite{itensor} and TeNPy~\cite{tenpy} libraries. In both cases, we use MPS bond dimensions up to $\chi_\textrm{max}=4000$ to ensure truncation errors in the final MPS of $\varepsilon_\textrm{trunc}\sim 10^{-5}$ at most. 

In order to find the QAH phase, which spontaneously breaks time-reversal symmetry, we add to the Hamiltonian a small complex field in each closed loop of NN, thus modifying the hopping strength $t$. That is, during the first sweeps of the DMRG algorithm, we add to the Hamiltonian a guiding field proportional to the QAH order parameter, i.e., of the form $ih\sum_{ij \in \textrm{plaq.}}  \hat{c}_i^\dagger\hat{c}_j$, for bonds $i\rightarrow j$ following the arrow convention of Fig.~\ref{fig:loop_order}, and with $h/t=10^{-2}$. With this procedure, one can target solutions with spatially inhomogeneous patterns of the QAH order parameter by imprinting such patterns in a spatially dependent guiding field. However, as the DMRG algorithm is in general not able to escape local minima induced by the spatial pattern of such a guiding field, it is crucial to compare the final energies of different initial patterns. For instance, the topological stripe state discussed in the main text can be stabilized with the DMRG algorithm at half filling, but it is a metastable state. That is, its energy is larger than the spatially inhomogeneous QAH state. On the contrary, the topological stripe state is the ground state of the system in the hole-doped case. 
\newpage

\section{Mean-field analysis of the Laughlin pump}
Here we analyze the Laughlin pump of our interacting model within the mean-field treatment. Even though the DMRG method is more accurate, the single-particle nature of the mean-field ansatz allows for a better understanding of the quantized transfer of charges, in terms of the spectral flow of mid-gap localized states. \\

For the mean-field analysis, we consider the same cylinder geometry and Hamiltonian of the main text, in which the free and interacting part read

\begin{equation} \label{eq:hamiltonianfree}
\begin{split}
\Hop_{0} = &-t\sum_{\langle ij\rangle}(\cdop_{i}\cop_{j}+\textrm{H.c.})+J\sum_{\llangle ij\rrangle}e^{i\phi_{ij}}(\cdop_{i}\cop_{j}+\textrm{H.c.}),
\end{split}
\end{equation}
and
\begin{equation} \label{eq:hamiltonianinter_p}
\begin{split}
\Hop_\textrm{int}= V_1\sum_{\langle ij\rangle} \hat{n}'_{i}\hat{n}'_{j}+V_2\sum_{\llangle ij\rrangle}\hat{n}'_{i}\hat{n}'_{j}
+V_3\sum_{\lllangle ij \rrrangle}\hat{n}'_{i}\hat{n}'_{j},
\end{split}
\end{equation}
with $\hat{n}'_{i}\equiv \hat{n}_{i}-1/2$ and $n_i=c^\dagger_i c_i$. We perform a standard Hartree-Fock decoupling of the density-density interaction terms
\begin{equation} 
\hat{n}_i\hat{n}_j \simeq -\xi_{ij}\cdop_j\cop_i-\xi_{ij}^*\cdop_i\cop_j+\abs{\xi_{ij}}^2+\bar{n}_i\nop_j+\bar{n}_j\nop_i-\bar{n}_i\bar{n}_j,
\label{eq:wicks}
\end{equation}
with $\xi_{ij}\equiv\langle\cdop_i\cop_j\rangle$ and $\bar{n}_i\equiv\langle\nop_i\rangle$. Notice that, within this approximation, the Hamiltonian becomes quadratic in the creation/annihilation fermionic operators. The ground state can then be found by iteratively diagonalizing the Hamiltonian with a Bogoliubov transformation, in order to determine the mean-field values $\xi_{ij}$ and $\bar{n}_i$ in a self-consistent loop. From previous studies~\cite{julia-farre_self-trapped_2020,cardarelli2022}, it is known that such mean-field ansatz can qualitatively capture the QAH phase of the system, with a shift in the interaction parameters in which the QAH phase appears, compared to the DMRG method. Therefore, here we change the interaction values used in the main text to $V_1=2.5t$, $V_2=1.5t$, and $V_3=0$, which leads to a QAH phase at half filling in the mean-field ansatz.\\
\subsection{Homogeneous QAH phase}
\begin{figure}[h]
    \centering
    \includegraphics[width=1\columnwidth]{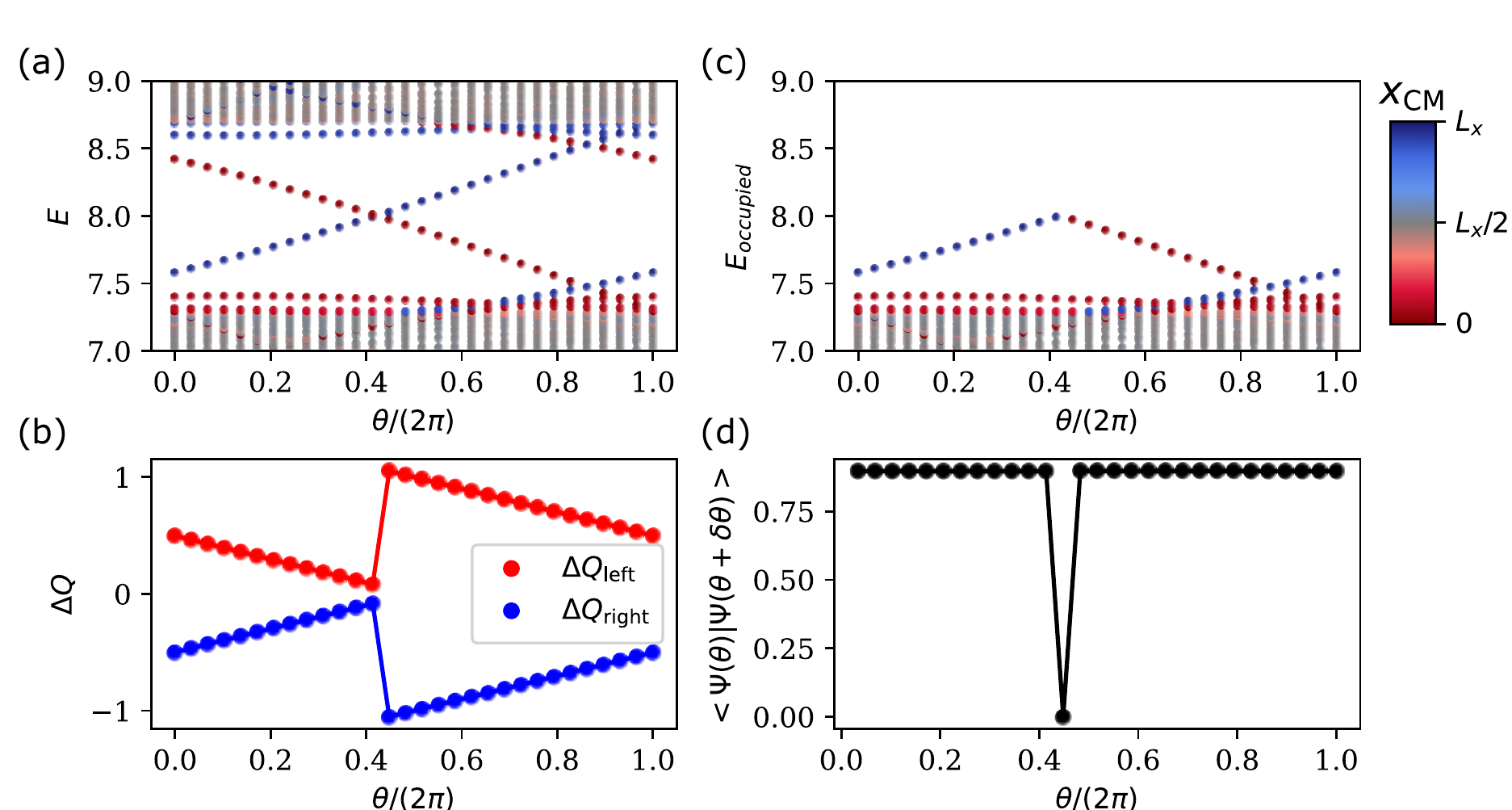}
    \caption{Mean-field results for a topological Laughlin pump in a cylinder with a spatially homogeneous interaction-induced QAH phase. The different quantities are plotted as a function of the inserted flux $\theta$. (a) Single-particle mean-field spectrum. Here the color code indicates the center-of-mass position of each state. (b) The net charge on each half of the cylinder. (c) Same as in (a) but showing only the populated states. (d) Overlap between consecutive wavefunctions.}
    \label{fig:simple_pump}
\end{figure}

For simplicity, let us start by analyzing the Laughlin pump in the homogeneous QAH phase found at half filling. We fix the mean-field parameters found self-consistently at the flux $\theta=0$, and investigate the response of the system under the flux insertion, as shown in Fig.~\ref{fig:simple_pump}. Figure~\ref{fig:simple_pump}(a) shows the mean-field single-particle energies of $\hat{H}(\theta)$ as a function of the inserted flux in the cylinder $\theta$, with the color code indicating their center-of-mass position. Notice that the spectrum of $\hat{H}(0)=\hat{H}(2\pi)$ is the same, and therefore the spectral flow of the left and right localized mid-gap states needs to be compensated by the bulk states. This leads to the charge flow from left to right, related to the presence of a nontrivial Chern number, which is shown in Figure~\ref{fig:simple_pump}(b). In contrast with the DMRG results, here we observe discrete jumps in the quantities $\Delta Q(\theta)$, resulting in $\Delta Q(2\pi)-\Delta Q(0)=0$ for both left and right edges. That is, even though the integrated slope of these quantities is quantized to $\pm 1$, no net charge transfer is observed. This is due to the breaking of adiabaticity in the single-particle method that we use, as shown in Figs.~\ref{fig:simple_pump}(c)-(d). Figure~\ref{fig:simple_pump}(c) shows the populated single-particle states during the pump. At the level crossing, and due to the fact that only half of the states are occupied at half filling, the right-localized edge state becomes unpopulated when crossing the Fermi energy, and the left-localized edge states get populated instead. This leads to a vanishing overlap between consecutive wavefunctions, $\ket{\Psi(\theta)},\ket{\Psi(\theta+\delta\theta)}$, as can be seen in Fig.~\ref{fig:simple_pump}(d). Indeed, one can rigorously relate the value of these jumps to the topological invariant, as discussed in Ref.~\cite{Hatsugai2016}. It is worth noting that, in the DMRG results presented in the main text, such jumps are absent due to the local nature of the state optimization within this method, and the fact that at each value of the flux $\theta$ we initialize the DMRG algorithm with the previous converged state at $\theta -\delta \theta$. Interestingly, this local nature of the state update resembles the situation that one would encounter in a real adiabatic time evolution under a flux insertion. Alternatively, for random initial states, we also observe jumps with the DMRG method (not shown here).

\newpage
\subsection{Topological stripe state}
\begin{figure}[h!]
    \centering
    \includegraphics[width=1\columnwidth]{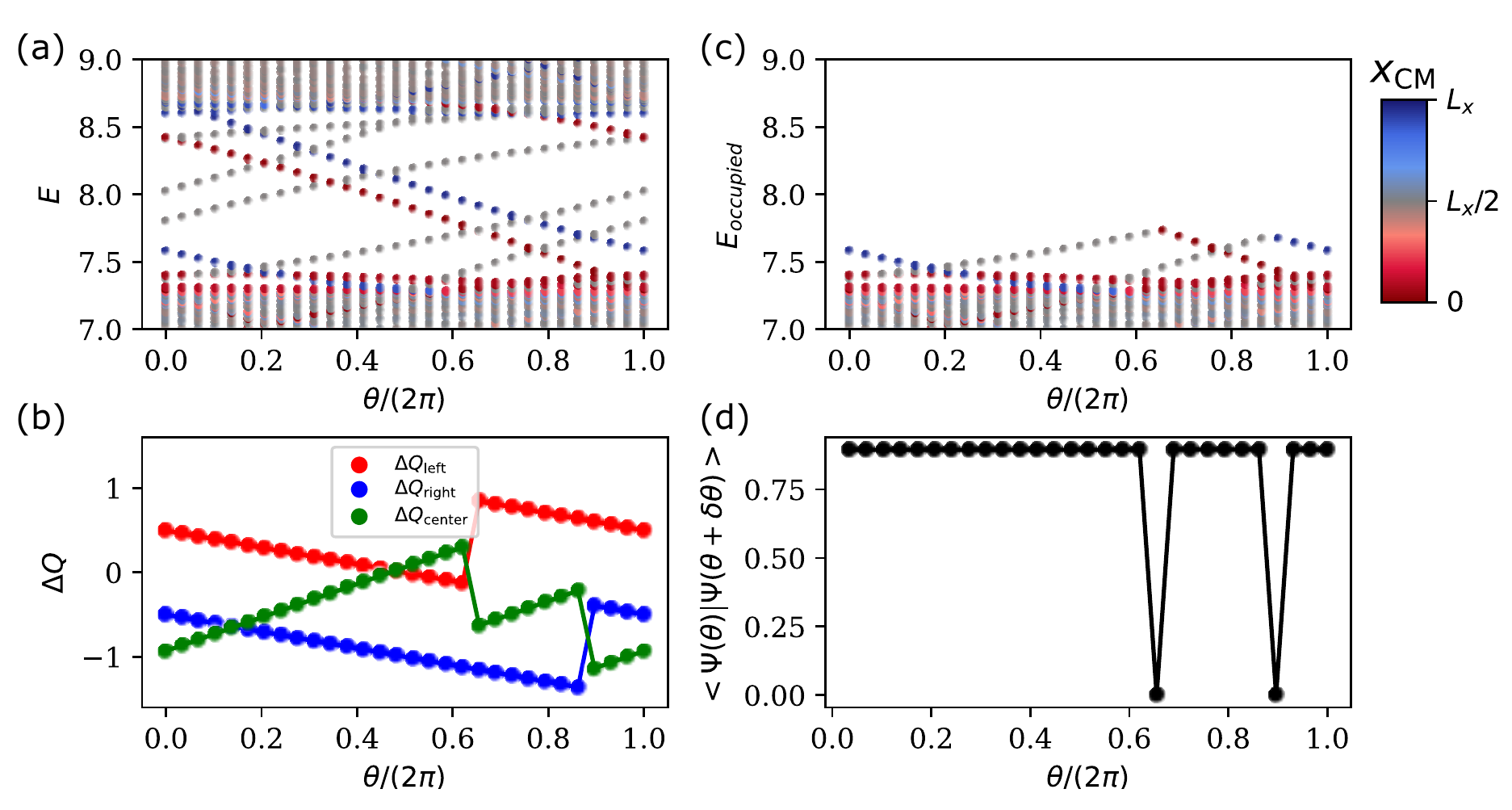}
    \caption{Mean-field results for a topological Laughlin pump in a cylinder with a spatially inhomogeneous interaction-induced QAH phase (topological stripe state). The different quantities are plotted as a function of the inserted flux $\theta$. (a) Single-particle mean-field spectrum. Here the color code indicates the center-of-mass position of each state. (b) The net charge on the left, center, and right regions of the cylinder. (c) Same as in (a) but showing only the populated states. (d) Overlap between consecutive wavefunctions.}
    \label{fig:domains_pump}
\end{figure}

The mean-field Laughlin pump in the topological stripe state can be considered as a generalization of the results presented in Fig.~\ref{fig:simple_pump}. In accordance with the DMRG results presented in the main text, the mean-field method in the cylinder geometry also leads to a domain wall pattern of the QAH phase for a single hole added to half filling. The response of this system to a periodic flux insertion is shown in Fig.~\ref{fig:domains_pump}. Figure~\ref{fig:domains_pump}(a) shows the appearance of additional midgap states, compared to the case of a homogeneous QAH phase of the previous Fig.~\ref{fig:simple_pump}. Such additional states are localized at the center domain wall of the cylinder and are a direct signature of the difference in the Chern number between the left and right halves of the cylinder.  This difference in Chern numbers also leads to an opposite direction of the charge flow on each half of the cylinder, as shown in Fig.~\ref{fig:domains_pump}(b), where one can observe that the charge flows from the left and right edges to the center, leading to a net accumulation in the center of the system. As in the homogeneous QAH phase, here we also observe jumps in the transferred charges, which are caused by the level crossings and vanishing overlaps between consecutive wavefunctions [see Figs.~\ref{fig:domains_pump}(c)-(d)].
\newpage

\bibliographystyle{apsrev4-1}
\bibliography{Biblio_1d}